\documentclass[12pt]{article}
\usepackage{graphicx}

\begin{document}

\title
{The Renormalization Effects in the Microstrip-SQUID Amplifier}
\author{G.P. Berman $^{1}$\footnote{Corresponding
author: gpb@lanl.gov}, A.A. Chumak$^{1,2}$, and V.I. Tsifrinovich$^{3}$
\\[3mm]$^1$
 Theoretical Division, MS-B213, Los Alamos National Laboratory, \\ Los Alamos, NM 87545
\\[5mm] $^2$  Institute of Physics of
the National Academy of Sciences,\\ pr. Nauki 46, Kiev-28, MSP 03028,
Ukraine
\\[3mm]$^3$
 Department of Applied Physics, Polytechnic Institute of NYU,\\ 6 MetroTech Center, Brooklyn, NY 11201
\\[3mm]\rule{0cm}{1pt}}
\maketitle \markright{right_head}{LA-UR 11-06109}

\begin{abstract}
The peculiarities of the microstrip-DC SQUID amplifier caused by the
resonant structure of the input circuit are analyzed. It is shown
that the mutual inductance, that couples the input circuit and the SQUID
loop, depends on the frequency of electromagnetic field. The
renormalization  of the SQUID parameters due to the screening
effect of the input circuit vanishes when the Josephson frequency is
much greater than the signal frequency.
\end{abstract}

\section{Introduction}

For a long time, the SQUID has been used as the most sensitive detector
of a magnetic flux. Recently, there has been growing interest in the
development of a low noise radio-frequency and microwave amplifiers,
for example, for axion detection \cite{bra,asz} or for the measurement
of superconducting quantum bits \cite{mic,ser}. For these
applications, the SQUID is a leading candidate due to its low power
dissipation and excellent noise properties.

In  typical SQUID amplifiers, the input signal is injected into an
input coil that is coupled to the SQUID washer. The input coil is
deposited on top of a dielectric layer, which covers the washer. The
coupling of the input circuit to the SQUID can significantly modify the properties
of both the SQUID and the input coil. A modification of the SQUID by a coupled
inductance was pointed out by Zimmerman \cite{zim} and studied by
Clarke and coworkers \cite{mar}-\cite{hil2}. The most important
influence of the input coil on the SQUID is the reduction of the
loop inductance. The macroscopic parameters of the SQUID take the
renormalized values corresponding to the reduced loop inductance.

Martinis and Clarke \cite{mar} have pointed out that the
renormalization effect, due to the mutual inductance of input circuit
and SQUID,  can be suppressed by parasitic capacitances between the
input coil and the washer. Parasitic capacitances, being widely
distributed, cause the gain to fall at high frequencies. More recently,
the idea of using the capacitance between the coil and the washer to
form a resonant microstrip has arisen \cite{muc} and developed
\cite{muc1}. The deleterious effect of the parasitic capacitance was
addressed by operating the input coil  as a transmission
line resonator. In this case, the input signal was applied between
one end of the coil and the washer, while the other end of the coil
was left open.

The transmission line resonator has an infinite set of
eigenfrequencies that is in contrast to a resonator with lumped
capacitance and inductance. In our model, for each eigenfrequency, $\omega_n$, we
will put in correspondence a couple of  values of $C_n$
and $L_n$, which satisfy the relation, $\omega_n=\big (L_nC_n\big
)^{-1/2}$. The input circuit response to the SQUID signals depends on
the characteristic frequencies of the circuit. The Josephson frequency,
$\omega_J$, and the frequency
 of the input signal, $\omega$, are the most important ones.
 Usually, the amplifier operates in the low frequency regime of
$\omega_J>>\omega$. Therefore, one can expect that the response of
the input circuit to the high-frequency SQUID signal will differ
significantly from the low-frequency response. The physical picture
is very similar to the effect of parasitic capacitances on SQUID
amplification.

\section{Transmission-line modes}

A non-dissipative lumped-element resonator with the inductance,
$L_{lump}$, and capacitance, $C_{lump}$, can be described by the
oscillator Hamiltonian,
\begin{equation}\label{one}
H=\frac {p^2}{2L_{lump}}+\frac {q^2}{2C_{lump}},
\end{equation}
where $L_{lump}$ and $(C_{lump})^{-1}$ are the effective mass and the spring
constant of the oscillator, and $p$ and $q$ are the corresponding momentum
and coordinate. The oscillator frequency is:
$\omega_r=(L_{lump}C_{lump})^{-1/2}$.

The superconducting transmission line is characterized by infinite number of
oscillators. Its Hamiltonian is \cite{el1,int}:
\begin{equation}\label{two}
H=\sum_n\frac 12\bigg(\frac {\pi^2_n}{l}+\frac
{\varphi^2_n}{c}k_n^2\bigg),
\end{equation}
where $\pi_n$ and $\varphi_n$ are the canonically conjugated ``momentum"
and ``coordinate",  $l$ and $c$ are the inductance and the
capacitance per unit length, respectively. In the case of open ends
of the line, the value of $k_n$ is equal to $\pi n/\Lambda$, where
$n=1,2,3,...$,  and $\Lambda$ is the length of the line. It follows from
Eq. (\ref{two}) that the frequency of the $n$th oscillator is,
\begin{equation}\label{thr}
\omega_n=k_n(lc)^{-1/2}.
\end{equation}
For each resonator mode, $n$, the current and the voltage profiles on
the resonator are given by sinusoidal and cosinusoidal
distributions, respectively,
\begin{equation}\label{vol}
I_n\sim\sin(n\pi x/\Lambda),\quad\;V_n\sim\cos(n\pi x/\Lambda),
\end{equation}
where the beginning of the line is at $x=0$.

After changing the variables, \[\pi_n^\prime=\pi_n/{\sqrt
{k_n}}~,\quad\,\varphi_n^\prime=\varphi_n{\sqrt {k_n}}~,\] Eq.
(\ref{two}) can be rewritten as:
\begin{equation}\label{fou}
H=\sum_n\frac 12\bigg(\frac {\pi^{\prime 2}_n}{l/k_n}+\frac
{\varphi^{\prime 2}_n}{c/k_n}\bigg).
\end{equation}
The quantities, $l/k_n$ and $\:c/k_n$, have the dimensions of the inductance and
capacitance, respectively. Comparing each summand in Eq. (\ref{fou})
with Eq. (\ref{one}), we conclude that the quantities, $L_n\equiv l/k_n$
and $C_n\equiv c/k_n$, are the inductance and the capacitance of the
$n$th resonator. The value, $(L_nC_n)^{-1/2}$, coincides with the
frequency, $\omega_n$, given by Eq. (\ref{thr}).

It follows that the values, $L_n$ and $C_n$,
depend on the number $n$ as $1/n$. Hence, $L_n=L_1/n,\:C_n=C_1/n$.
For the case of the fundamental frequency ($n=1$), we have:
\begin{equation}\label{fiv}
L_1= {{\Lambda}\over{\pi}}l,~C_1={{\Lambda}\over{\pi}}c.
\end{equation}

A set of resonances in the transmission line can modify the
renormalization  of the SQUID parameters. In what follows we
analyze this phenomenon in more details.

\section{Scheme of the microstrip-SQUID amplifier}

Fig. 1 shows schematically the equivalent circuit of the
microstrip-SQUID amplifier. $V_i$ is the amplitude of the input voltage;
$R_i$ and $R$ are the resistances of the voltage source and the
resistance shunting the stripline, respectively; $C_i$ is the
coupling capacitance.

\begin{figure} [ht]
\centering
\includegraphics{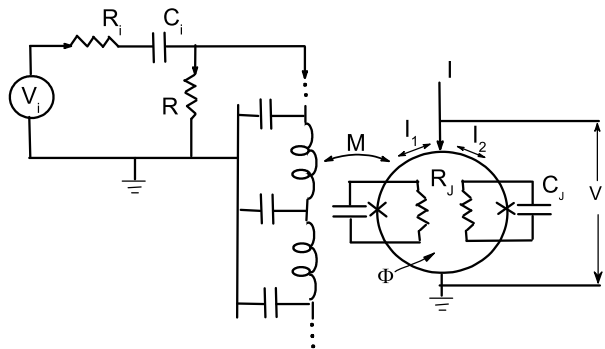}
\caption{The equivalent circuit of the microstrip-SQUID amplifier. The
resonant microstrip is shown as a sequence of lumped inductances and
capacitances. The mutual inductance, $M$, couples the input circuit
to the SQUID loop whose inductance is, $L_J$.}
\end{figure}

The two Josephson junctions are in parallel with the resistances, $R_J$
(which can be external shunts) and capacitances, $C_J$. The shunting
effect of these elements provides nonhysteretic characteristics for
the SQUID. The Josephson junctions are connected in parallel, thus
forming a superconducting loop of inductance, $L_J$. The transmission
line is coupled to the SQUID loop by the mutual inductance, $M$. An
external dc current bias, $I$, and a dc flux, $\Phi$, are applied to
provide the most favorable values of the transfer
function, ${dV}/{d\Phi}\equiv V_\Phi$ , where $\Phi$ is the magnetic
flux inside the loop. Here we will not take into account the noise
component of the electromagnetic fields which is present in the
circuit due to the resistances, $R_J,\:R,\:R_i$, and generated by
external sources.
\begin{figure} [ht]
\centering
\includegraphics{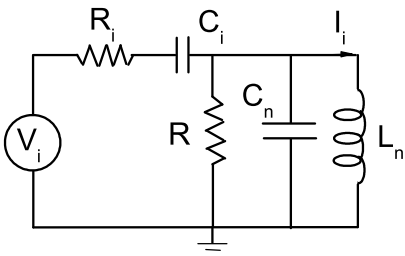}
 \caption{Lumped element model of the input circuit. In the vicinity
 of the $n$th resonance ($\omega\approx \omega_n$) the transmission
 line is represented  by the capacitance, $C_n$,
 and the inductance, $L_n$.}
\end{figure}

The transmission line current caused by the input voltage, $V_i$,
depends on the frequency, $\omega$. In the vicinity of some
eigenfrequency, $\omega_n$, the input circuit can be described by
the equivalent scheme shown in Fig. 2.  Usually $\omega$ is close to
the fundamental frequency, $\omega_1$. For this particular case and
in the absence of the SQUID, the  forward impedance of the input
circuit, defined as $Z= {V_i}/{I_i}_{|M=0}$,
 is:
\begin{equation} \label{six}
Z=\frac {L_1}{C_i}\bigg\{i\bigg[\omega\bigg(C_1+C_i+C_i\frac
{R_i}R\bigg)-\frac 1{\omega L_1}\bigg]+\frac
1R-R_iC_i\bigg(\omega^2C_1-\frac 1{L_1}\bigg)\bigg\}.
\end{equation}
As we see, the resonant properties of the input circuit depend not
only on the parameters of the microstrip but also on the presence of
other elements. (The effect of coupling capacitance
on amplification was studied experimentally in \cite{dar}.) Also,
the SQUID parameters affect the input circuit characteristics via
the mutual inductance, $M$.

Similar to $L_n$ and $C_n$, the value of $M$ depends on the
frequency of radiation. The following  can help to determine
this dependence. It can be seen from Eq. (\ref{vol}), that the
current of the $n$th mode changes its direction at $x=\Lambda m/n$,
where $m$ is an integer ($m<n$). The total flux generated by the
$n$th mode in the SQUID loop is smaller than it could be in the case
of a constant sign (for example, in the case of $n=1$). It is
reasonable to approximate $M(\omega_n)$ by
\begin{equation}\label{sev}
M(\omega_n)\equiv M_n=M_1/n.
\end{equation}
 The coefficient $1/n$ accounts for the reduction of the high-frequency
 flux, generated in the loop. Eq.(\ref{sev}) is a good approximation
 for a signal frequency,
$\omega\approx \omega_1$, as well as for the Josephson frequency,
$\omega>> \omega_1$. In what follows, both characteristic
frequencies are of our major interest.

For high frequencies, where discreteness of the transmission line
spectrum is not important, we can consider
\begin{equation}\label{eig}
M(\omega)\approx M_1\frac {\omega _1}\omega .
\end{equation}
The theoretical analysis of amplification should account for the
dependence of $M$ on the frequency of electromagnetic field.

\section{Equations of motion for the coupled microstrip-SQUID}

In this Section, we analyze how the frequency dependence of $M$,
given by Eq. (\ref{eig}), modifies the renormalization effect of the SQUID
parameters. We will see that the renormalization vanishes if the
Josephson frequency is much larger than the input signal frequency.

The standard system of equations describing the dynamics of the SQUID is
given by:
\[J=\frac 12(I_2-I_1),\]
\[I=I_1+I_2,\]
\[V=\frac {\Phi_0}{4\pi}\bigg(\dot{\delta}_1+\dot{\delta}_2\bigg),\]
\[\Phi+L_JJ+\widetilde{MI_i}=\frac {\Phi_0}{2\pi}\bigg(\delta_1-\delta_2\bigg),\]
\[\frac {\Phi_0}{2\pi}C_J\ddot{\delta}_1+\frac{\Phi_0}{2\pi R_J}\dot{\delta}_1=
I_1-I_c\sin\delta_1, \]
\begin{equation}\label{nin}
\frac {\Phi_0}{2\pi}C_J\ddot{\delta}_2+\frac{\Phi_0}{2\pi
R_J}\dot{\delta}_2= I_2-I_c\sin\delta_2 ,
\end{equation}
where, $\delta _{1,2}$, is the phase difference across the left or
right junction, respectively, $\Phi _0=\frac h{2e}$ is the flux
quantum, $e$ is the electron charge ($e>0$ here), $I_c$ is the
critical value of the Josephson current through an individual
junction. A wide tilde, ($\widetilde{...}$), indicates the convolution:
\begin{equation}\label{ten}
 \widetilde{MI_i}=\int_{-\infty}^t dt^\prime M(t-t^\prime)I_i
(t^\prime),
\end{equation}
where the mutual inductance, $M(t)$, ($M(t)=0$ for $t<0$) can
 be expressed via its
 Fourier-transform as,
\[M(t)=\frac 1{2\pi}\int_{-\infty}^{+\infty}d\omega e^{i\omega
t}M(\omega).\]

The fourth equation in (\ref{nin}) explicitly describes  the
SQUID-input circuit coupling via the mutual inductance, $M$. This
coupling arises from the current, $I_i$, which induces a magnetic flux
in the SQUID loop, thus changing the voltage, $V$, across the SQUID.

On the other hand, the current around the loop, $J$, induces a
voltage, $-\partial_ t \widetilde{MJ}$, in the transmission line of the input
circuit. Hence, the  current, $I_i$, is generated by both the
external voltage, $V_i$, and the voltage caused by the circulating
current, $J$. In the frequency domain, $I_i$ is given by:
\begin{equation}\label{Ele}
 I_i(\omega^\prime)=\frac{V_i(\omega^\prime)}{Z_{\omega^\prime}}-i
 \frac {\omega^\prime M(\omega^\prime)J(\omega^\prime)}
 {Z_{\omega^\prime}}\frac {C(\omega^\prime)}{C_i},
\end{equation}
where  the input voltage is considered to be a harmonic function,
$V_i(t)=V_ie^{i\omega t}$, and $V_i(\omega^\prime)=2\pi V_i\delta
(\omega-\omega^\prime)$. We have also assumed, for simplicity, that the resistance of
the source and of the coupling capacitance are small: $R_i<<R,\:C_i<<C$.

The quantities, $L$ and $C$, depend on the frequency, $\omega^\prime$,
similar to the dependence of $M(\omega^\prime)$:
\begin{equation}\label{Twe}
L(\omega^\prime)\approx L_1\frac {\omega _1}{\omega^\prime}
,\:\,C(\omega^\prime)\approx C_1\frac {\omega _1}{\omega^\prime }.
\end{equation}

Substituting the expression (\ref{Ele}) for $I_i$   into the fourth
equation of the system (\ref{nin}), we have:
\[\Phi+L_JJ+\frac {M_1V_i(t)}{Z(\omega)}-\]
\begin{equation}\label{Thi}
\int_{-\infty}^tdt^\prime
M(t-t^\prime)\int_{-\infty}^{+\infty}\frac{d\omega^\prime}{2\pi}
i\omega^\prime e^{i\omega^\prime t^\prime}\frac{M_{\omega^\prime}
J_{\omega^\prime}C_{\omega^\prime}}{Z(\omega^\prime)C_i}=
 \frac{\Phi_0}{2\pi}\bigg(\delta_1-\delta_2\bigg).
\end{equation}
The third term in the left-hand side of Eq. (\ref{Thi}) describes the
low-frequency (the input voltage frequency) contribution to the flux
threading the SQUID loop. The fourth term represents both the
low-frequency and the high-frequency components. The high
frequency is in the  range close to $\omega_J$. It follows from the
 analysis of Section 3 (see Eq. (\ref{eig})) that,
\begin{equation}\label{Fou}
 M^2(\omega _J)
 \approx M_1^2\frac {\omega_1^2}{\omega_J^2}<<M_1^2.
 \end{equation}
Eq. (\ref{Fou}) shows that the high-frequency contribution to the
total flux arising from strip-line-SQUID coupling can be neglected.
In this case, Eq. (\ref{Thi})  reduces to
\begin{equation}\label{Fif}
\Phi+L_JJ+\Delta\Phi(t)=\frac
{\Phi_0}{2\pi}\bigg(\delta_1-\delta_2\bigg),
\end{equation}
where $\Delta\Phi (t)$  contains only the low-frequency
Fourier-components:
\begin{equation}\label{Six}
\Delta \Phi(\omega^\prime)=\frac
{M_1V_i(\omega^\prime)}{Z(\omega^\prime)}-\frac{C_1M_1^2}{C_i}
 i\omega^\prime\frac {
J_{\omega^\prime}} {Z(\omega^\prime)}.
\end{equation}
The frequency, $\omega^\prime$, in Eq. (\ref{Six}) is in the
vicinity of the input voltage frequency $\omega$
($|\omega^\prime-\omega|<<\omega )$.

Following the arguments of Ref. \cite{mar}, we assume that the
low-frequency components, $J(\omega)$ and $V(\omega )$, are
generated by a small input voltage, $V_i$. By this reason, we consider
both of them to be small quantities also. Considering a solution of the
unperturbed SQUID equations to be known, we can express the linear
response to the ``external" flux, $\Delta\Phi$, as
\begin{equation}\label{Sev}
V(t)=V_\Phi \Delta \Phi(t),\;\;J(t)=J_\Phi \Delta \Phi(t),
\end{equation}
where $V_\Phi \equiv \partial V/\partial \Phi$ and $J_\Phi \equiv
\partial J/\partial \Phi$. In contrast to the main result of the
Reference \cite{mar}, the derivatives in Eq. (\ref{Sev}) should be
those of the bare SQUID circuit (with non-renormalized inductance,
$L_J$). This means that the high-frequency screening  of the input
circuit is negligible.

Using Eqs. (\ref{Fif})-(\ref{Sev}), we can easily obtain the
microstrip amplifier gain. If we represent the output voltage as
$V(t)=Ve^{i\omega t}$, then the gain is given by:
\begin{equation}\label{Eig}
\frac {V}{V_i}=\frac {M_1V_\Phi}{Z_\omega +J_\Phi i\omega
M_1^2C_1/C_i} .
\end{equation}
The second term in the denominator of Eq. (\ref{Eig}) describes the
modification of the low-frequency input impedance. This modification
is due to the SQUID back action. At the same time, the coefficients
$V_\Phi$ and $J_\Phi$ are taken for a bare SQUID. This corresponds
to the physical picture in which the effective coupling of both
sub-systems through the mutual inductance occurs at the signal
frequency. The coupling becomes weak at high (Josephson) frequency.

\section{Conclusion}
Using the transmission line as a resonator of the input
circuit does not result in a renormalization of the SQUID parameters, if the
Josephson frequency is much greater than the frequency of the input
signal. A similar effect was analyzed in \cite{mar}. The authors
have discussed the influence of parasitic capacitances between the
turns of the input coil or between the input coil and the SQUID, on the
amplifier dynamics. It was shown that there was no current flow at
the Josephson frequency in the input circuit in the presence of
large parasitic capacitances. In this case, the reduced SQUID
parameters are replaced with the bare SQUID parameters.

The presence or absence of the renormalization effect is of great
importance for optimization of the input circuit parameters and
achieving desirable gains especially  in the gigahertz frequency
region. It is worth mentioning, that the maximum SQUID amplification
is restricted by the value of $R_J/L_J$ \cite{joh}. The renormalized
inductance of the loop is smaller than the bare inductance, $L_J$.
This explains the tendency of gain decrease with the increase of the
signal frequency that was observed
 experimentally (see, for example, Ref. \cite{muc3}).

Summarizing, in the limit of very high Josephson frequencies, there
is no difference what kind of capacitance is in the input circuit.
In both cases (parasitic irregular capacitance or smoothly
distributed capacitance of the transmission line), a similar
shunting effect causes the reflection of the SQUID signal by the
input circuit. This prevents a renormalization of the SQUID
parameters.

\section{Acknowledgment}

We thank D. Kinion  for useful discussions.  This work was carried out under the auspices of the National Nuclear Security Administration of the
U.S. Department of Energy at the Los Alamos National Laboratory under Contract No. DE-AC52-
06NA25396, and was funded by the Office of the Director of National Intelligence (ODNI), and Intelligence Advanced Research Projects Activity (IARPA). All statements of fact, opinion or conclusions contained herein are those of the authors and should not be construed as representing the official views or policies of IARPA, the ODNI, or the U.S. Government.

\newpage \parindent 0 cm \parskip=5mm

\end{document}